\newcommand{\pasp}{Publ. Astron. Soc. Pac.}

\newcommand{\mnras}{Mon. Not. R. Astron. Soc.}
\newcommand{\apjl}{Astrophys. J. Lett.}
\newcommand{\apjs}{Astrophys. J. Suppl. Ser.}
\newcommand{\apj}{Astrophys. J.}
\newcommand{\aj}{Astron. J.}

\newcommand{\aap}{Astron. Astrophys.}
\newcommand{\aaps}{Astron. Astrophys. Suppl.}

\newcommand{\pasa}{Publ. Astron. Soc. Aust.}

\newcommand{\aapr}{Astron. Astrophys. Rev.}

\newcommand{\an}{Astron. Nachr.}

\documentclass[universe,article,accept,moreauthors]{Definitions/mdpi} 

\usepackage{comment}
\usepackage{gensymb}
\graphicspath{{fig/}}
\firstpage{1} 
\pubvolume{1}
\issuenum{1}
\articlenumber{0}
\pubyear{2026}
\copyrightyear{2026}
\externaleditor{Firstname Lastname} 
\datereceived{29 June 2026} 
\daterevised{7 August 2026} 
\dateaccepted{9 August 2026} 
\datepublished{ } 

\Title{The Total and Polarized Radio Emission from the Innermost~Jets~of a High-Redshift Quasar and a Candidate at~Parsec-Scale~Resolution}

\Author{Bence Husv\'eth $^{1}$, Krisztina \'E. Gab\'anyi $^{1,2,3,}$*\orcidA{}, Tao An $^{4,5,}$*\orcidB{}, S\'andor Frey $^{1,3}$\orcidC{}, Jun Yang $^{6}$\orcidD{}, Iv\'an Agudo  $^{7}$\orcidE{} \mbox{and Yingkang Zhang  $^{5}$\orcidF{}}}

\AuthorNames{Bence Husv\'eth, Krisztina \'E. Gab\'anyi, Tao An, S\'andor Frey, Jun Yang, Iv\'an Agudo and Yingkang Zhang }

\address{%
$^{1}$ \quad Department of Astronomy, Institute of Physics and Astronomy, ELTE Eötvös Loránd University,  Pázmány Péter sétány 1/A, 1117 Budapest, Hungary\\
$^{2}$ \quad HUN-REN--ELTE Extragalactic Astrophysics Research Group, ELTE Eötvös Loránd University, Pázmány Péter sétány 1/A, 1117 Budapest, Hungary\\
$^{3}$ \quad Konkoly Observatory, HUN-REN Research Centre for Astronomy and Earth Sciences, MTA Centre of Excellence, Konkoly Thege Miklós út 15-17, 1121 Budapest, Hungary\\
$^{4}$ \quad Department of Astronomy, University of Science and Technology of China, Hefei 230026, China\\ 
$^{5}$ \quad Key Laboratory of Radio Astronomy, Shanghai Astronomical Observatory, Chinese Academy of Sciences, \mbox{80 Nandan Road}, Shanghai 200030, China\\
$^{6}$ \quad Department of Physics and Astronomy, Chalmers University of Technology, Onsala Space Observatory,  \mbox{43992 Onsala, Sweden}\\
$^{7}$ \quad Instituto de Astrof\'isica de Andaluc\'ia, IAA-CSIC, Glorieta de la Astronom\'ia s/n, 18008 Granada, Spain

} 

\corres{Correspondence: k.gabanyi@astro.elte.hu (K.É.G.); antao2008@ustc.edu.cn (T.A.)}

\abstract{High-frequency very long baseline interferometry (VLBI) polarimetry probes synchrotron-emitting plasma closer to the central engines of radio-loud active galactic nuclei (AGNs), but observations above $43$\,GHz are technically demanding. We present $22$-GHz European VLBI Network observations of the $z=4.31$ quasar J1510$+$5702 and J1606$+$3124, whose published spectroscopic redshift, $z=4.56$, is uncertain; a photometric estimate gives $z_{\rm phot}=0.9\pm0.1$. For the published $z>4$ redshifts, $22$\,GHz corresponds to rest-frame frequencies above $118$\,GHz. Polarized emission is detected in J1510$+$5702, and a low-level polarized signal is recovered from the brightest feature of J1606$+$3124. Adopting $z=4.56$, that feature has a brightness temperature of $T_{\mathrm{b,VLBI}}=(7.4\pm0.8)\cdot10^{10}$\,K, allowing a mildly Doppler-boosted interpretation, while the young compact-source scenario also remains viable. The core of J1510$+$5702 has $T_{\mathrm{b,VLBI}}=(1.08\pm0.15) \cdot10^{12}$\,K, implying a Doppler factor of ${\sim}22$ under the equipartition assumption. This component has a ${\sim}3.5$\% fractional polarization. These observations show that cm-wavelength VLBI can access rest-frame millimeter-band polarization in bright $z>4$ jets.}


\begin{document} 


\section{Introduction}
Active galactic nuclei (AGNs) are extremely powerful, continuously radiating sources that emit across the electromagnetic spectrum, from radio waves to $\gamma$-rays \cite{Padovani}. The energy required for this comes from accretion onto the central supermassive black hole (SMBH; $M \sim 10^6\text{--}10^9\,M_{\odot}$). 
In some cases, relativistic outflows (jets) can be observed perpendicular to the accretion disk. These are the so-called radio-loud or jetted AGNs, which, however, only account for ${\lesssim}10$\% of the AGN population \cite{jet10prc,2025ApJS..280...23A}. 

\textls[-15]{The radio emission from the jets has a non-thermal origin; it is generated by synchrotron radiation from charged particles moving at relativistic speeds in magnetic fields. Synchrotron radiation is inherently polarized, and jet polarization therefore traces the ordered component of the magnetic field and the magneto-ionic material along the line of sight \cite{Pacholczyk1970,Agudo2014}.}

If the jet is inclined at a small angle to the line of sight ($i\lesssim10^\circ$), relativistic effects, such as Doppler boosting, become important. The approaching jet emission is enhanced, while the counter-jet is deboosted by a similar factor, producing the apparent one-sided jet morphology typically observed in so-called blazar-type radio-loud AGNs \cite{agn_fig}. On the other hand, if the jets are oriented closer to the plane of the sky, no strong relativistic beaming is expected. 
In such a case, if a symmetric, double-lobed radio structure is observed with a steep spectrum, and the overall extent of the entire radio morphology does not exceed a linear size of ${\sim}1$\,kpc, then the source can be classified as a compact symmetric object (CSO)~\cite{1994ApJ...432L..87W,Readhead2021}. These objects are often gigahertz-peaked spectrum (GPS) sources, showing a flux-density peak at a frequency of ${\sim}1$~GHz in their radio spectrum \cite{ODea2021}. They are characterized by slow variability (${<}20\%\;\mathrm{yr}^{-1}$) and low apparent jet speeds (${<}2.5\,c$, where $c$ denotes the speed of light) \cite{CSOI, Readhead2021, An2025}. Their sizes are smaller than their host galaxies~\cite{fanti,fanti2}. Their compactness is traditionally explained by two models, the `youth scenario' and the `frustration scenario'. The former one states that these objects are newborn jetted sources, representing the earliest state of a radio galaxy evolution, and after sufficient time, they can grow into Fanaroff--Riley type radio galaxies \cite{readhead94cso,size_freq,anb2012}. On the other hand, according to the `frustration scenario', the jets may be progressing in a dense interstellar medium (ISM), which prevents their growth~\cite{oDea91,vanBreugel84,An2026}.

Since both blazars and CSOs are compact radio sources, mapping the jet morphology requires imaging observations with high angular resolution. It is possible with the very long baseline interferometry (VLBI) technique, allowing for resolution at milliarcsecond (mas) scales at GHz observing frequencies.

Powerful AGNs can be observed from vast cosmological distances; accreting SMBHs are now identified out to $z \approx 10$ \cite{z10}, and the highest-redshift known blazars lie at \mbox{$z \approx 6\text{--}7$ \cite{Spingola_blazar,2025NatAs...9..293B}}. Due to the expansion of the Universe, cm-wavelength observations of high-redshift sources probe their emission at $(1+z)$ times higher rest-frame frequencies, or equivalently at $(1+z)$ times shorter rest-frame wavelengths. In blazar jets, progressively higher radio frequencies probe regions closer to the central engine because of the frequency-dependent opacity of synchrotron-emitting plasma \cite{1979ApJ...232...34B,1998A&A...330...79L}. At mm wavelengths, the innermost compact regions of the jets can be observed; these regions can be optically thick and therefore obscured at longer wavelengths (e.g., \cite{Lee2008}). The effects of Faraday rotation and depolarization are also less severe at higher radio frequencies \cite{Agudo2014}. A useful benchmark for high-redshift VLBI polarimetry is J0906$+$6930 at $z=5.47$ \cite{an20b}. VLBI observations performed at $15$~GHz, corresponding to a rest-frame frequency of $97$~GHz, revealed polarized emission at a position slightly displaced from the jet core, tracing an apparent bending in the jet and indicating interaction between a nascent jet and the dense surrounding interstellar medium \cite{an20b}. 

In this paper, we present mas-scale, polarization-sensitive VLBI observations taken at $22$\, GHz of CGRaBS\,J1606$+$3124 and ICRF\,J151002.9$+$570243 (hereafter J1606$+$3124 and J1510$+$5702, respectively). J1510$+$5702 has a secure spectroscopic redshift of $z=4.31$~\cite{J1510z} and is one of the highest-redshift quasars detected in $\gamma$-rays (e.g., \cite{gokus2022,Benke}). It also hosts a ${\sim}15$-kpc X-ray jet discovered with {\it Chandra} \cite{Siemiginowska,yuan}. J1606$+$3124 can instead be regarded as a candidate high-redshift AGN. Its cataloged spectroscopic redshift, \mbox{$z=4.56$}, was flagged as uncertain by \cite{R_flux}, while the DESI Legacy Imaging Surveys DR8 photometric estimate gives $z_{\rm phot}=0.9\pm0.1$ \cite{desi}. It has been classified as both a gigahertz-peaked-spectrum \mbox{galaxy~\cite{oDea91,Stanghellini93}} and a flat-spectrum radio quasar \cite{R_flux}. Simultaneous multi-frequency radio observations showed a peaked spectrum with a turnover frequency at \mbox{$(1.8\pm0.1)$~GHz~\cite{mingalev, Sotnikova21}}. Unless stated otherwise, we adopt the published value of $z=4.56$ to permit direct comparison with earlier VLBI work, but every rest-frame frequency, brightness temperature, linear size, and apparent speed quoted for J1606$+$3124 is conditional on that redshift. We examine the $z=0.9$ alternative explicitly in Section\,\ref{sec:J1606_disc}.

In the following, we assume a flat $\Lambda$CDM cosmological model where the value of the Hubble constant is $H_0=70\,\mathrm{km\,s^{-1}\, Mpc^{-1}}$, the matter density parameter is $\Omega_\mathrm{m}=0.27$, and the vacuum energy density parameter is $\Omega_{\Lambda}=0.73$. At $z=4.56$ and $z=4.31$, $1$\,mas corresponds to projected linear scales of $6.789$\,pc and $6.962$\,pc, respectively \cite{cosmocalc}. We define the spectral index, $\alpha$, as $S \propto \nu^\alpha$, where $S$ is the flux density and $\nu$ the observing frequency. The position angles are measured from north through east.

\section{Observations and Data Reduction} \label{sec:obs}

The VLBI observations were carried out in June $2018$ at a central frequency of $22$~GHz with the European VLBI Network (EVN) (project code: EZ028, PI: Tao An). The same frequency setup was used for both targets: eight intermediate frequency (IF) bands, each with a bandwidth of $16$\,MHz. All four Stokes products were recorded. The correlation was performed at the Joint Institute for VLBI European Research Infrastructure Consortium (JIVE, Dwingeloo, The Netherlands) with the SFXC Software Correlator \cite{corr2015} with a correlation integration time of $2$\,s. Further details of the observations and the target sources are listed in Table
~\ref{obs}. 

\begin{table}[H] 
\caption{Details of the $22$-GHz EVN observations. The columns give the name of the source, observing date, on-source integration time in seconds, adopted redshift, and the luminosity distance in Mpc~\cite{cosmocalc}.\label{obs}}
\begin{tabularx}{\textwidth}{CcCCC}
\toprule 
\textbf{Name}	& \textbf{Obs. Date}	& \textbf{\boldmath{$t$} (s)} &\boldmath{$z$} &\textbf{\boldmath{$ÍD_\mathrm{L}$} (Mpc)}\\
\midrule
J1606$+$3124	& 10 June 2018 & 23,251 & 4.56\textsuperscript{a} & 43,292.2\textsuperscript{a}\\
J1510$+$5702	& 13 June 2018 & 34,062 & 4.31 & 40,489.7 \\ 
\bottomrule
\end{tabularx}
    \noindent\footnotesize{\textsuperscript{a} The DESI DR8 gives a photometric redshift estimate as $z_{\rm phot}=0.9\pm0.1$ \cite{desi}. The luminosity distance corresponding to this lower redshift value is ${\sim}5900$\,Mpc \cite{cosmocalc}.}
\end{table}

\begin{figure}[H]
\centering
\begin{adjustwidth}{-\extralength}{0cm}
\subfloat[\centering J1606$+$3124]{\includegraphics[width=8.55cm]{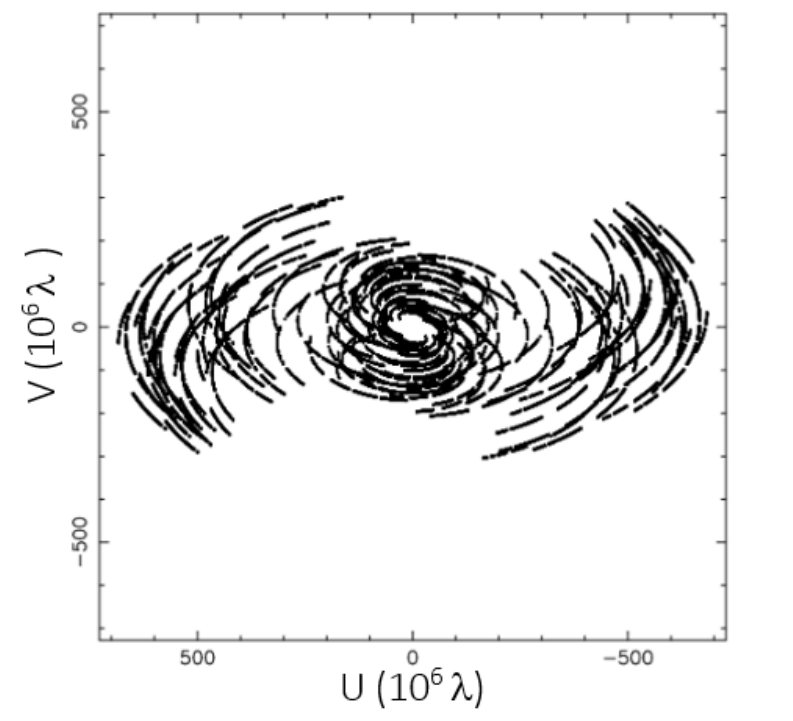}}
\hfill
\subfloat[\centering J1510$+$5702]{\includegraphics[width=8cm]{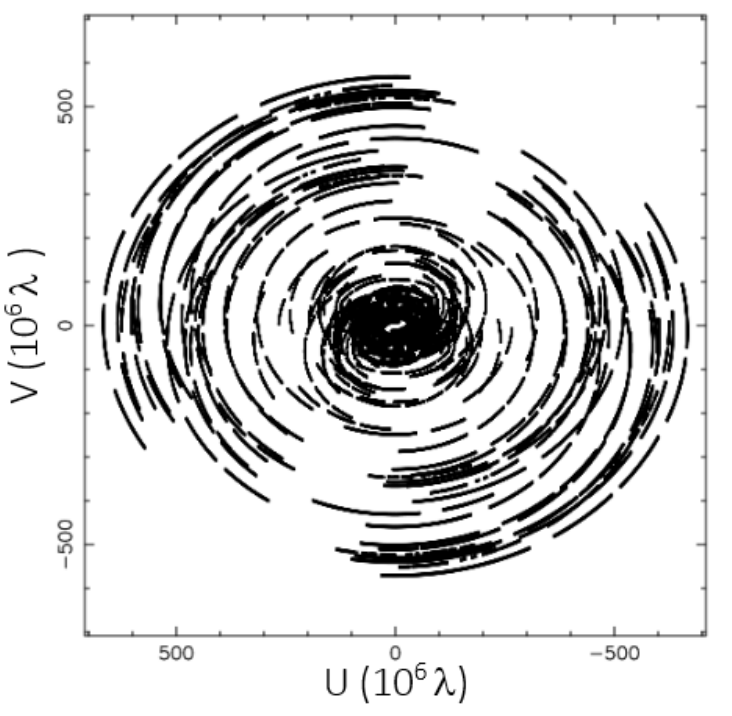}}
\end{adjustwidth}
\caption{The $(u,v)$-coverages of the observations carried out at $22$ GHz with the EVN for J1606$+$3124~(\textbf{a}) and J1510$+$5702 (\textbf{b}). The axes give the $u$ and $v$ coordinates of the baseline vectors in units of million wavelengths ($10^6 \lambda$), where $\lambda = 1.35$~cm. \label{uv}}
\end{figure}

In the observation of J1606$+$3124,
$13$ EVN telescopes participated: Effelsberg (EF, Germany), Jodrell Bank Mrk 2 (JB, UK), Medicina (MC, Italy), Noto (NT, Italy), Sardinia (SR, Italy), Onsala (O6, Sweden), Toru\'n (TR, Poland), Yebes (YS, Spain), Svetloe (SV, Russia), Zelenchukskaya (ZC, Russia), Badary (BD, Russia), Yonsei (KY, Republic of Korea) and Tamna (KT, Republic of Korea). All but one of the telescopes participated in the observation of J1510$+$5702. KT was not able to observe J1510$+$5702 due to technical difficulties. The obtained $(u,v)$-coverages (i.e., distributions of baseline vectors projected onto the plane perpendicular to the source direction) for the two measurements are displayed in Figure~\ref{uv}.

Besides the two target sources, the following calibrators were observed: 3C\,345, NGC\,6251, J1638$+$5720, and J1551$+$5806 together with J1606$+$3124, and 3C\,454.3 and 3C\,84 together with J1510$+$5702. Additionally, J2007$+$777 was used as a fringe finder in both observations.

\subsection{Total Intensity}

The data were reduced in the National Radio Astronomy Observatory (NRAO) Astronomical Image Processing System ({\sc aips} \cite{aips}) following standard VLBI procedures. The a priori amplitude calibration used the gain and system-temperature information supplied by the EVN pipeline, and parallactic-angle corrections were applied. We used the provided flag tables to remove erroneous data points. Additionally, we edited out the 7th IF of SR in the observation of J1510$+$5702 because it was severely affected by radio frequency interference, as indicated in the information letter received from the correlator. EF was used as the reference antenna in both observations.

After the initial instrumental phase delay corrections, global fringe-fitting was performed for all calibrators. The resulting data were channel-averaged and exported to {\sc difmap} \cite{difmap}. We produced hybrid images of the calibrators through several iterations of {\sc clean} \cite{Hogbom} and phase-only self-calibration. Antenna- and IF-dependent multiplicative gain corrections were then derived from the calibrators with the best $(u,v)$ coverage, J1638$+$5720 and 3C\,454.3, and transferred in {\sc aips} to all sources, including the targets. Afterwards, fringe-fitting was done for the target sources, and their total intensity data were imaged in {\sc difmap} involving several iterations of {\sc clean} and phase-only self-calibration. Amplitude self-calibration was applied only after the source model could no longer be improved by phase self-calibration alone. The solution time intervals were gradually decreased from a length comparable to the whole on-source observing time down to one~minute.

To quantitatively describe the brightness distributions of the sources, the calibrated visibility data were fitted with two-dimensional Gaussian components in {\sc difmap}. The uncertainties of the component widths and positions were calculated with the formulas from \cite{Kun, Schinzel}. The flux density errors were calculated based on \cite{Fomalont}, with $10\%$ uncertainty added in quadrature, as commonly adopted for EVN measurements at this frequency, to account for VLBI absolute flux density calibration uncertainties.

\subsection{Polarization Calibration} \label{sec:polcal}

To perform right- and left-hand delay corrections needed for polarization calibration, we used the \texttt{VLBACPOL} procedure in {\sc aips} after the instrumental delay corrections but before fringe-fitting. We also verified the delay corrections with the task \texttt{RLDLY}; we did not find significant differences between the two approaches. We used the calibrators J1638$+$5720 and 3C\,454.3 to derive the delay corrections. After fringe-fitting, we derived the D-terms (polarization leakage terms) using the {\sc aips} task \texttt{LPCAL}, using J1638$+$5720 and 3C454.3 for the observations of J1606$+$3124 and J1510$+$5702, respectively. For 3C\,454.3, we defined two polarized components with a distance of ${\sim}0.5$\,mas in the brightest region of the source, following the results of the Monitoring of Jets in Active Galactic Nuclei with Very Long Baseline Array (VLBA) Experiments (MOJAVE \cite{mojave_pol}) at $15$~GHz and of the VLBA-BU-BLAZAR program  ({\url{https://www.bu.edu/blazars/BEAM-ME.html} accessed on \mbox{15 March 2026}}) at $43$~GHz (e.g., \cite{bostonblazar}). We self-calibrated the non-channel-averaged total intensity data of the calibrators in {\sc difmap}. The five edge channels at both ends of each IF were discarded because of their lower sensitivity. The {\sc aips} task \texttt{IMAGR} was used to clean the self-calibrated visibility data of the calibrators separately for Stokes parameters Q and U, after applying the D-term corrections. The resulting Q and U images were combined with the task \texttt{COMB} to create polarized-intensity and polarization-angle images.
 
 Before transferring the leakage solutions to the targets, we inspected the complex D-terms antenna by antenna, separately for the two circular hands and for all usable IFs. Table~\ref{tab:dterms} reports the median amplitudes, full antenna ranges, and the median IF-to-IF rms scatter. We also repeated \texttt{LPCAL} with an independent calibrator in each experiment, where its parallactic-angle coverage and polarized structure permitted a stable solution. These sources were 3C345 and 3C84 for J1606$+$3124 and J1510$+$502, respectively. For 3C345, we defined two polarized components, while we regarded 3C84 as unpolarized. The median absolute differences between the leakage terms are given in the last column of Table\,\ref{tab:dterms}. 

\begin{table}[H]
\caption{Diagnostics of the instrumental-polarization solutions. The medians and ranges are calculated over the participating antennas after excluding flagged data. $\sigma_{\rm IF}$ is the median antenna-by-antenna rms scatter among the usable IFs. The last column gives the median absolute complex-D-term differences obtained with independent leakage calibrators. These were 3C345 and 3C84 in the case of J1606$+$3124 and J1510$+$5702, respectively.\label{tab:dterms}}
\begin{adjustwidth}{-\extralength}{0cm}
\begin{tabularx}{\linewidth}{CCCCCCC}
\toprule
\textbf{Target} & \textbf{Adopted Calibrator} & \textbf{Median \boldmath{$|D_R|$}} & \textbf{Median \boldmath{$|D_L|$}} & \textbf{Full \boldmath{$|D|$} Range} & \boldmath{$\sigma_{\rm IF}$} & \textbf{Independent-Solution Difference}\\
 & & \textbf{(\%)} & \textbf{(\%)} & \textbf{(\%)} & \textbf{(Percentage Points)} & \textbf{(Percentage Points)}\\
\midrule
J1606$+$3124 & J1638$+$5720 & $5.5$ & $6.8$ & 0.7--47.8 & $7.3$ & $8.6$\\
J1510$+$5702 & 3C\,454.3 & $4.9$ & $6.2$ & 0.3--29.0 & $6.3$ & $8.0$\\
\bottomrule
\end{tabularx}
\end{adjustwidth}
\end{table}

The derived D-term corrections were copied over to the target sources and applied in the same way as for the calibrators during the cleaning of the self-calibrated data of Stokes Q and U with \texttt{IMAGR}. Similarly, the \texttt{COMB} task was used to combine the Q and U images to create the polarized intensity and polarization angle images. To further assess the quality of the polarization calibration, we performed these steps with the alternative D-terms as well. For the polarized flux densities quoted below, the Ricean bias correction (e.g., \cite{george2012}) is negligible at the measured signal-to-noise ratios (${\gtrsim}7$ for J1606$+$3124 and ${\gtrsim}15$ for J1510$+$5702); applying the standard debiasing formula changes the integrated polarized flux densities by less than a few percent.

We compared the resulting Stokes Q and U images of the target sources obtained with the adopted calibrator solutions and with the alternative solutions. We calculated the differences between the polarized intensities as 
\begin{equation}
    \Delta P= \sqrt{(Q_1-Q_2)^2 + (U_1-U_2)^2},
\end{equation}
where $Q_1$ and $Q_2$, and $U_1$ and $U_2$ are the Stokes Q and U flux densities measured within the areas where the highest polarized intensities were detected in the images obtained with the adopted or the alternative D-term calibrations. In the case of J1510$+$5702, the difference was modest, $0.4$\,mJy compared with the polarized intensity values; thus, we added this difference in quadrature to the thermal noise when reporting the polarized flux density value. For J1606$+$3124, the polarized intensity value obtained with the alternative \mbox{D-term} calibrations is ${\sim}60$\,\% higher than the one obtained by using the calibrator J1638$+$5720. This is most probably related to the inadequate modeling of the radio structure of 3C\,345 (due to the poorer $(u,v)$-coverage), which has a complex polarized substructure (e.g., \cite{Roder_2024}).

To determine the electric vector position angles (EVPAs), one would need calibrated polarization observations of the polarized calibrator performed close in time and at the same frequency. No such strictly contemporaneous $22$-GHz observations of our polarized calibrators were available to us. However, 3C\,454.3 was observed relatively close in time, on 22 April 2018 at $15$~GHz within the MOJAVE \cite{mojave_pol} and 16 June 2018 at $43$~GHz within the VLBA-BU-BLAZAR program \cite{bostonblazar}. At both frequencies, the EVPAs of the jet component were similar, ${\sim}90\degree$. This is consistent with the negligible rotation measure (RM), \mbox{$(-4 \pm 74)$\,rad\,m$^{-2}$}, derived for this component between $8$~GHz and $15$~GHz on \mbox{15 June 2006} by \cite{mojave_pol}. We therefore assumed that the EVPA of this jet feature at $22$\,GHz was also close to $90\degree$ and used this assumption for the absolute EVPA calibration of J1510$+$5702. Because this calibration is not based on a strictly contemporaneous $22$-GHz measurement, the absolute EVPAs should be regarded as approximate. We therefore use them below only for qualitative comparison with earlier measurements, and not for an RM determination. We also note that \cite{mojave_pol} reported RM variability in 3C 454.3; for example, $(-136 \pm 72)$\,rad\,m$^{-2}$ was derived for an observation on 9 March 2006.

\textls[-15]{The closest polarized VLBI measurement of the other calibrator, J1638$+$5720, that we are aware of was performed at $15$~GHz within the MOJAVE program on \mbox{19 August 2018 \cite{mojave2023}} with a reported EVPA of $93\degree$ ({\url{https://www.cv.nrao.edu/MOJAVE/} accessed on \mbox{24 June 2026}}). The formal extrapolation using the RM value $352.7 \pm 79$\,rad\,m$^{-2}$ derived by the MOJAVE team \cite{mojave_pol} from measurements taken on 24 May 2006 would change the EVPA between $15$ and $22$\,GHz by only ${\sim}4\degree$. Nevertheless, since the higher-frequency EVPA behavior of J1638$+$5720 is not established by contemporaneous data, and since intrinsic EVPA variability could dominate over this formal RM extrapolation, we decided not to use this information. Thus, we did not perform an absolute EVPA calibration for J1606$+$3124.}

\section{Results}

\subsection{Total Intensity}

\begin{figure}[H]
\centering
\includegraphics[width=14cm, bb=80 170 700 515, clip]{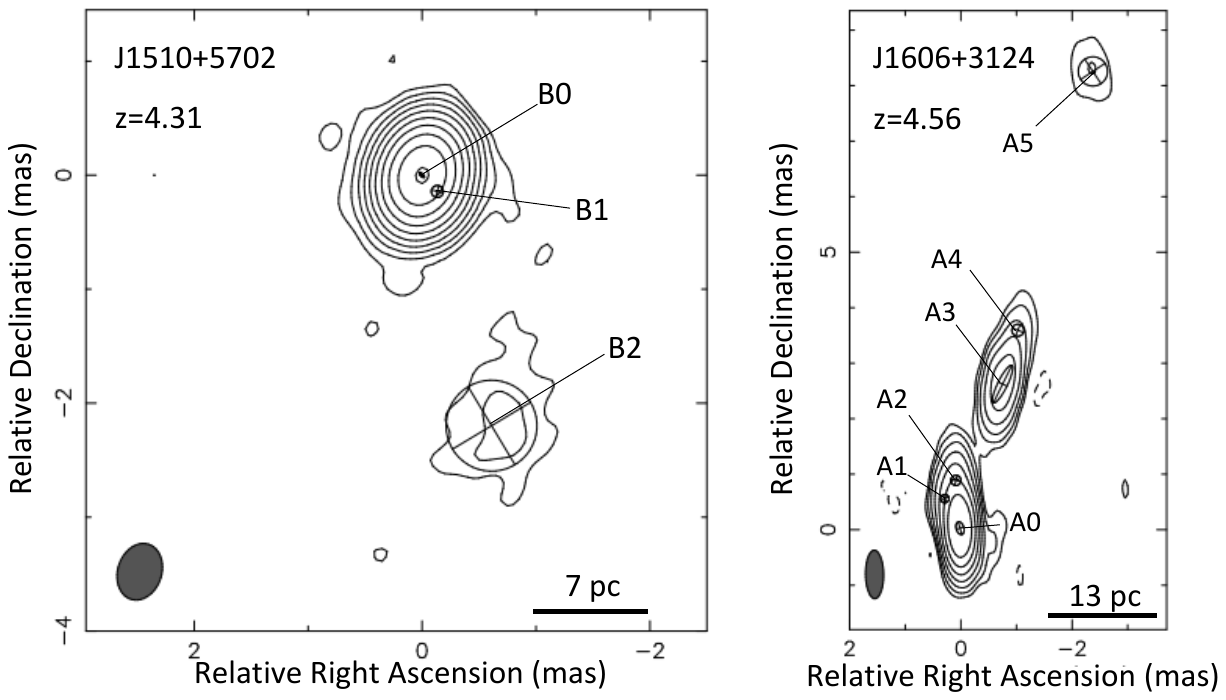}
\caption{EVN images at $22$\,GHz of the two targets made from the self-calibrated visibility data, with the fitted two-dimensional Gaussian model components overlaid. The positions and full widths at half maximum (FWHM) of the components are indicated. The gray ellipses in the lower left corners represent the FWHM of the restoring beam. In the lower right corners, the linear scales are displayed.    
\textit{Left panel:} The map of the quasar J1510$+$5702. The peak intensity is $102.7\;\mathrm{mJy\;beam^{-1}}$, and the lowest contours are drawn at $\pm 0.2\;\mathrm{mJy\;beam^{-1}}$, corresponding to an image noise level of $\pm 3\sigma$. The contour values increase by a factor of two. The FWHM of the restoring beam is $0.5\;\mathrm{mas}\,\times\;0.4\,\mathrm{mas}$, and the position angle of the major axis is $\mathrm{PA}=-18\degree$. 
\textit{Right panel:} The map of the quasar J1606$+$3124. The peak intensity is $152\,\mathrm{mJy\,beam^{-1}}$, and the lowest contours are drawn at $\pm 0.9\;\mathrm{mJy\;beam^{-1}}$, corresponding to an image noise level of $\pm 4\sigma$. 
The FWHM of the restoring beam is $0.9\;\mathrm{mas}\,\times\;0.3\,\mathrm{mas}$, with $\mathrm{PA}=1\degree$.
\label{map_comp}}
\end{figure}

The naturally weighted $22$-GHz EVN intensity map of J1606$+$3124 is displayed in the right panel of Figure~\ref{map_comp}. It shows a roughly north--south-oriented structure with a slight turn of a few degrees towards the west at ${\sim}2$\,mas from the brightest feature, corresponding to a ${\sim}14$\,pc projected linear distance at the redshift of the source. The whole extent of the radio structure is ${\sim}9$\,mas (${\sim}60$\,pc projected linear size). 

\textls[-15]{The visibilities could be equally well described by either six or seven Gaussian components. The additional seventh component could be placed on the southwestern side of component A0. However, its inclusion does not significantly affect the parameters of the other components or the rms (root-mean-square) value of the obtained residual map. Additionally, it is pointlike with an unrealistically small size, most probably an artifact related to the sparse visibility sampling. Therefore, we used six Gaussian components to describe the brightness distribution of J1606$+$3124; their parameters are listed in Table \ref{j16_param}. Except for A0 and A3, all fitted components have circular Gaussian brightness distributions. The northernmost component, A5, is at a position angle of ${\sim}$$-16^\circ$ from the brightest feature, A0.}

\vspace{-18pt}

\begin{table}[H] 
\caption{Parameters of the fitted brightness distribution model components of the quasar J1606$+$3124. Columns from 1 to 7 give the components' designations, relative coordinates in right ascension and declination directions with respect to the brightest component, the FWHM sizes of the major and minor axes, position angles of the major axes, and flux densities, respectively. Missing minor axes and position angles indicate a fitted circular Gaussian component. \label{j16_param}}
\begin{adjustwidth}{-\extralength}{0cm}
\begin{tabularx}{\linewidth}{CCCCCCC}
\toprule
\textbf{ID} & \textbf{\boldmath{$\Delta$}RA} & \textbf{\boldmath{$\Delta$}Dec.} & \boldmath{$W_{\mathrm{maj}}$} & \boldmath{$W_{\mathrm{min}}$} & \textbf{PA}  &\boldmath{$S$}\\
& \textbf{(mas)} & \textbf{(mas)} & \textbf{(mas)} & \textbf{(mas)} & \textbf{(\boldmath{$\degree$})} &  \textbf{(mJy)} \\
\midrule 
    A0 & 0 & 0 & $0.233 \pm 0.006$ & $0.139 \pm 0.006$ &  $17 \pm 3$ & $172.0 \pm 17.3$ \\     A1 &  $0.27 \pm 0.05$ & $0.5 \pm 0.1$ & $0.15 \pm 0.02$ & $\ldots$ & $\ldots$ & $11.7 \pm 1.4$ \\ 
    A2  & $0.07 \pm 0.05$ & $0.9 \pm 0.1$ & $0.18 \pm 0.02$ & $\ldots$ & $\ldots$ & $19.3\pm 2.1$ \\ %
    A3  & $-0.77 \pm 0.05$ & $2.6 \pm 0.1$ & $0.75 \pm 0.02$ & $0.17 \pm 0.02$ & $-26 \pm 3 $ & $44.0\pm 4.5$ \\ 
    A4  & $-1.04\pm 0.05$ & $3.6\pm0.1$ & $0.22\pm0.05$ & $\ldots$ & $\ldots$ & $3.2\pm 0.5$ \\ 
    A5 & $-2.39\pm 0.07$ & $8.2\pm 0.1$ & $0.5\pm0.1$ & $\ldots$ & $\ldots$ & $3.4\pm 0.6$ \\ 
\bottomrule
\end{tabularx}
\end{adjustwidth}
\end{table}

In the left panel of Figure~\ref{map_comp}, the $22$-GHz EVN map of J$1510+5702$ is shown. It consists of two main emitting regions, a brighter northern and a fainter southern one, at ${\sim}2$\,mas to the south-southwest, corresponding to a projected linear distance of ${\sim}14$\,pc in the rest frame of the source. The visibilities can be adequately fitted with three Gaussian components, two circular and one elliptical. The latter is the brightest, most compact component, B0. The parameters of the model components are listed in Table \ref{j15_param}.

\begin{table}[H] 
\caption{Parameters of the brightness distribution model components fitted to the visibilities of the quasar J1510$+$5702. The column descriptions are the same as in Table\,\ref{j16_param}.
 \label{j15_param}}
\begin{adjustwidth}{-\extralength}{0cm}
\begin{tabularx}{\linewidth}{CCCCCCC}
\toprule
\textbf{Component} & \textbf{\boldmath{$\Delta$}RA} & \textbf{\boldmath{$\Delta$}Dec.} & \boldmath{$W_{\mathrm{maj}}$} & \boldmath{$W_{\mathrm{min}}$} & \textbf{PA} &\boldmath{$S$}\\
& \textbf{(mas)} & \textbf{(mas)} & \textbf{(mas)} & \textbf{(mas)} & \textbf{(\boldmath{$\degree$})} & \textbf{(mJy)} \\
\midrule 
    B0  & $0$ & $0$ & $0.055\pm 0.002$ & $0.023\pm 0.002$ & $48 \pm 3$ & $102.0\pm10.2$ \\ 
    B1 & $-0.13\pm0.05$ & $-0.14\pm0.07$ & $0.106\pm0.002$& $\ldots$ & $\ldots$ & $2.9\pm0.3$ \\ 
    B2 & $-0.6\pm0.1$ & $-2.2\pm0.1$ & $0.8\pm0.1$ & $\ldots$ & $\ldots$ & $1.4\pm0.3$ \\ 
\bottomrule
\end{tabularx}
\end{adjustwidth}
\end{table}

For both sources, all fitted components are larger than the smallest resolvable sizes with the interferometer in these observations, as determined in \cite{Kovalev}. We calculated the brightness temperatures of the components for both sources following the formula in \cite{Veres}:
\begin{linenomath}
\begin{equation}
T_{\mathrm{b,VLBI}}=1.22\cdot10^{12}(1+z)\frac{S}{W_{\mathrm{maj}}\cdot W_{\mathrm{min}}\cdot\nu^2} \mathrm{\;\;(K)},
\label{T_b}
\end{equation}
\end{linenomath}
where $z$ is the redshift of the quasar, $S$ the flux density of the component in Jy, $W_{\mathrm{maj}}$ and $W_{\mathrm{min}}$ are the FWHM sizes in mas of the major and minor axes, respectively, and $\nu$ the central observing frequency in GHz. For circular Gaussian components, the tabulated FWHM was used for both axes. Assuming that the intrinsic brightness temperature equals the equipartition value, 
 $T_{\mathrm{eq}}=5\cdot10^{10}\;\mathrm{K}$ \cite{Readhead94}, one can estimate the Doppler-boosting factor, $\delta$, as follows:
\begin{linenomath}
\begin{equation}
\delta=\frac{T_{\mathrm{b,VLBI}}}{T_{\mathrm{eq}}}.
\label{delta}
\end{equation}
\end{linenomath}

The brightest components of both quasars exceed the equipartition value, strongly in J1510$+$5702 and modestly in J1606$+$3124, suggesting relativistic boosting if the equipartition intrinsic brightness temperature is adopted. No other components had $T_{\mathrm{b,VLBI}}$ in excess of the equipartition value. The obtained $T_\mathrm{b, VLBI}$ and $\delta$ values are listed in Table \ref{j16_T} for components where relativistic boosting is possible. The listed uncertainties are propagated from the fitted component parameters and the adopted flux-density scale; they do not include the systematic uncertainty associated with the assumed intrinsic brightness temperature.

\begin{table}[H] 
\caption{
Brightness temperatures and Doppler-boosting factors (assuming the equipartition value for the intrinsic brightness temperature) calculated for the brightest, compact components of J1606$+$3124 and J1510$+$5702. The values derived for J1606$+$3124 adopt the uncertain redshift of $z=4.56$. If its redshift is $z_{\rm phot}=0.9$, then $T_{\rm b,VLBI}=2.5\cdot10^{10}$\,K and Doppler boosting is not required. The identification of A0 as the radio core of J1606$+$3124 is model-dependent (see text for details).
 \label{j16_T}}
\begin{tabularx}{\linewidth}{CCC}
\toprule 
 \textbf{Source} &
\boldmath{ $T_{\mathrm{b,VLBI}}$} \boldmath{$(10^{10}\;\mathrm{K})$} &
 \boldmath{$\delta$} \\
\midrule 
    J1606$+$3124---A0 & $7.4\pm 0.8$& $1.5\pm0.2$\\ 
    J1510$+$5702---B0 & $107.9\pm 14.8$ & $21.6\pm3.0$ \\ 
\bottomrule
\end{tabularx}
\end{table}

\subsection{Polarization}

\begin{figure}[H]
    \includegraphics[width=0.99\textwidth, bb=0 0 810 530, clip]{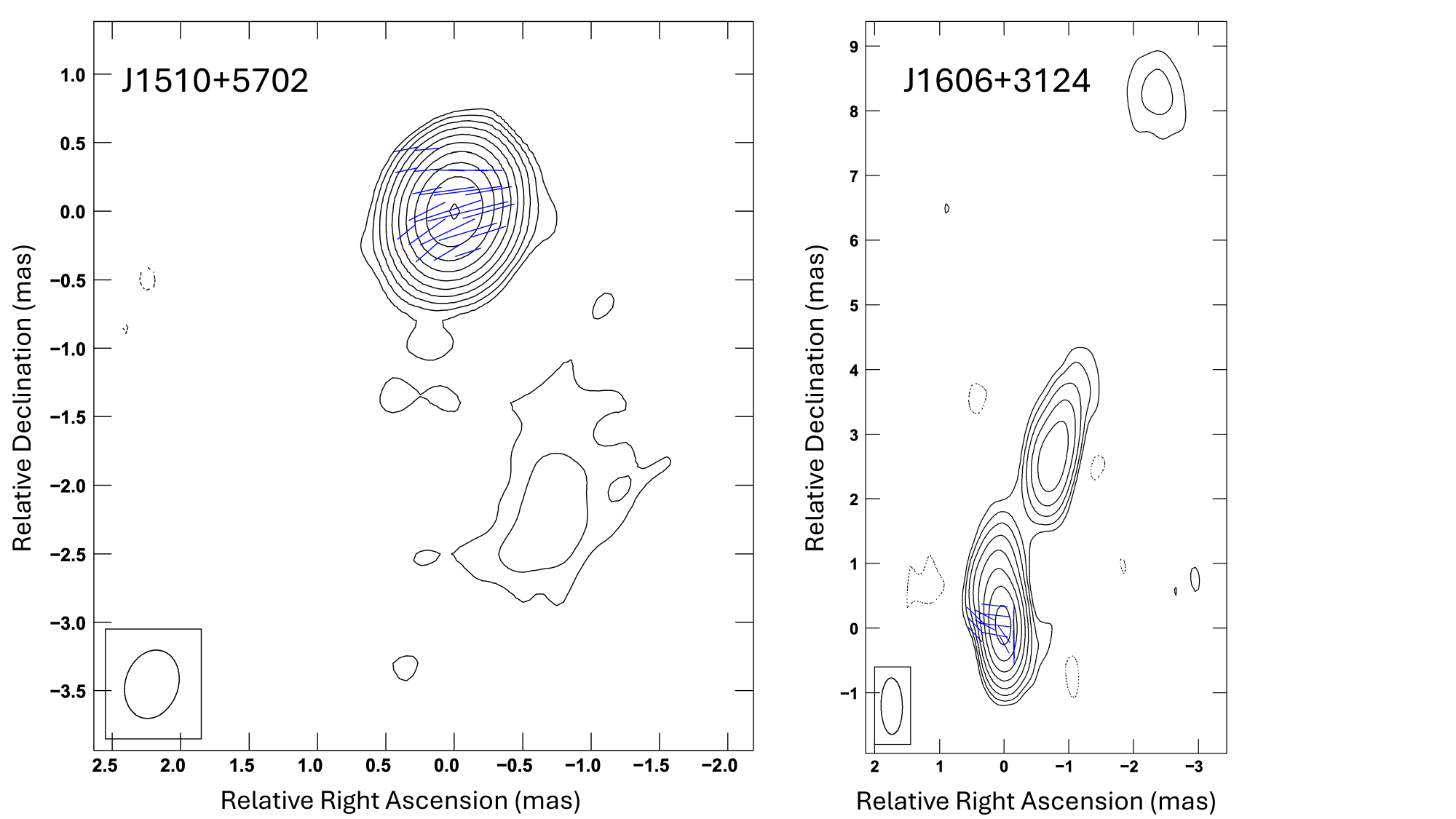}
    \caption{Total intensity (contours) and polarization (lines) images of J1510$+$5702 (\textbf{left}) and J1606$+$3124 (\textbf{right}) at $22$~GHz. The details of the total intensity images are the same as in Figure\,\ref{map_comp}. {\it Left panel: } The lines represent the direction of the EVPA after the approximate absolute calibration described in Section~\ref{sec:polcal}. A length of $0.1$\,mas corresponds to $1$\,mJy\,beam$^{-1}$ polarized intensity. {\it Right panel:} The lines represent the relative EVPA, but the absolute direction of EVPA is not calibrated. A length of $0.1$\,mas corresponds to $0.3$\,mJy\,beam$^{-1}$ polarized intensity.}
    \label{fig:pol}
\end{figure}

A low-level polarized signal was detected at the position of the brightest total intensity feature in J1606$+$3124 (Figure\,\ref{fig:pol}, right panel). The existence of the polarized signal is consistent irrespective of whether we used the D-term corrections derived from J1638$+$5720 or 3C345. However, the values of integrated polarized flux density are very different for the two D-term solutions (see Section\,\ref{sec:polcal}). The polarized flux density is in the range of $(2\text{--}3.2)$\,mJy. For a total flux density of ${\sim}144$\,mJy in the same region, the fractional polarization is ${\sim}(1.4\text{--}2)$\,\%. Due to the low level of polarized signal and the dependence of the obtained value on the D-term calibrator, we regard this detection as tentative only. No physical meaning is assigned to EVPA changes within one restoring beam. The absolute EVPA is not calibrated.

We detected polarization in the central region of J1510$+$5702 with a polarized flux density of {$p_{22}^\mathrm{J1510}=(3.4\pm0.4)\,\mathrm{mJy}$. The quoted error includes the thermal noise derived and propagated from the Stokes Q and U images, and the uncertainty derived by using the different D-term calibrators. The difference between the polarized flux density obtained using the D-term solutions from 3C\,454.3 and 3C\,84 is $0.4$\,mJy.} In the same region, the flux density in the total intensity image is {${\sim}95.8$\,mJy; thus, the fractional polarization is $3.5 \pm 0.5\%$.} The pixel with the highest polarized intensity is at the center with a value of ${\sim}3$\,mJy\,beam$^{-1}$. The {blue} polarization vectors together with the total intensity image are shown in the left panel of Figure \ref{fig:pol}. We rotated the obtained {EVPAs by $100\degree$ as inferred from 3C\,454.3 (Section~\ref{sec:polcal})}. The obtained mean EVPA of J1510$+$5702 is ${\sim}107\degree$ (equivalent to $-73\degree$ under the usual $180\degree$ EVPA ambiguity). This value should be treated as approximate because of the non-contemporaneous EVPA calibration. {We therefore refrain from interpreting small EVPA changes across the unresolved B0--B1 total intensity region.}

We did not detect polarization in the more extended, jet-like features of our target sources down to approximately $3\sigma$ polarized-intensity levels of $1.0$\,mJy\,beam$^{-1}$ and $0.7$\,mJy\,beam$^{-1}$ for J1606$+$3124 and J1510$+$5702, respectively.

\section{Discussion}

\subsection{J1606+3124} \label{sec:J1606_disc}

The $22$-GHz radio structure of J1606$+$3124 agrees well with that recovered at $8$~GHz by \cite{an2022}, while the higher-frequency data resolve the inner region in much finer detail. The two epochs are separated by $76$\,d in the observer frame (corresponding to ${\sim}14$\,d at $z=4.56$). The largest separation rate measured by \cite{an2022}, $0.013\pm0.002$\,mas\,yr$^{-1}$ for N relative to S, predicts a displacement of only $0.0027\pm0.0004$\,mas over $76$\,d. The C--S displacement is ${\sim}0.0012$\,mas. These shifts are negligible relative to the restoring beams and fitted component sizes, so component motion does not limit the cross-identification. The proper motions, however, do not constrain flux-density evolution. For that uncertainty, we use the directly measured $15$-GHz modulation index, $V=0.11$, from 2014.4--2020 \cite{an2022}. 
 Treating an $11\%$ flux-density change as a conservative characteristic inter-epoch uncertainty gives
$\sigma_{\alpha,{\rm var}}=\ln(1+0.11)/\ln(22.2/8.4)=0.11$.
This likely overestimates stochastic variability over only $76$\,d because the published light curve is dominated by a slow decline, but it provides a quantitative allowance for non-simultaneity. The different angular resolutions remain a separate systematic and prevent component-by-component spectral indices within blended regions.

{The outermost 22-GHz component A5 is associated with N in \cite{an2022}. Their component C corresponds to A3 + A4, while S + S0 corresponds to A0 + A1 + A2. The flux densities and the derived two-point spectral indices are listed in Table~\ref{tab:spectral-index}. Including the variability term in quadrature gives total uncertainties of $0.2$ for the spectral indices of the southern, middle, and northern regions. The spectral indices of the southern and middle regions are statistically indistinguishable, so their small difference cannot identify the core. The radio spectrum of the northern region is much steeper than either inner region even under the adopted variability allowance. Thus, the robust result is the strong outer-to-inner spectral steepening, but the derived spectral indices of the inner regions cannot be used as a unique core diagnostic.}

\begin{table}[H] 
\caption{Two-point spectral indices calculated from the non-simultaneous $8.4$- and $22$-GHz VLBI data for J1606$+$3124. The $8.4$-GHz component flux densities are from \cite{an2022}; the $22$-GHz values are from this work. The quoted uncertainty on $\alpha$ combines the propagated flux-density error and a variability contribution $\sigma_{\alpha,{\rm var}}=0.11$ estimated from the published $15$-GHz modulation index.}

    \label{tab:spectral-index}
\begin{tabularx}{\linewidth}{CCCCC}
\toprule 
 \textbf{ID\boldmath{$_{8.4}$}} & \textbf{\boldmath{$S_{8.4}$} (mJy)} & \textbf{ID\boldmath{$_{22}$}} & \textbf{\boldmath{$S_{22}$} (mJy)} & \boldmath{$\alpha$} \\
\midrule 
        S + S0 & $321\pm15$ & A0 + A1 + A2 & $203.0\pm17.5$ & $-0.5\pm0.2$ \\
        C & $82\pm4$ & A3 + A4 & $47.2\pm4.5$& $-0.6\pm0.2$\\
                N & $36\pm 2$ & A5 & $3.4\pm0.6$ & $-2.4\pm0.2$\\
\bottomrule
\end{tabularx}
\end{table}

An et al. (2022) \cite{an2022} described the object as a possible compact symmetric object (CSO) with the center of the AGN located in their component C, approximately at the position of our component A3. The steep radio spectrum they obtained for the northern and southern features using $2$- and $8$-GHz VLBI data and the overall peaked radio spectral shape of J1606$+$3124 led them to this classification. It was further supported by the slow long-term flux density variability and the mildly relativistic jet speeds they detected. In the new, higher angular resolution $22$-GHz observation, the southernmost, brightest component, A0, has $T_\mathrm{b,VLBI}$ modestly above the equipartition value, giving a nominal Doppler factor of $\delta\sim 1.5$. This modest excess alone is not sufficient to rule out a compact hot spot interpretation of A0, but it shows that weak Doppler boosting is allowed by the data. If this Doppler factor is combined with the intrinsic jet speed of $\beta=(0.6\text{--}0.8)$ inferred by \cite{an2022}, it implies an inclination angle of ${\sim}40^\circ$, consistent with the ${\gtrsim}28^\circ$ value given in \cite{an2022}. The detected one-sided polarization could also be explained within the CSO scenario, where the approaching, brighter jet side is more polarized due to a smaller amount of intervening depolarizing material (e.g., \cite{Tremblay2016}). 

Nevertheless, several properties keep a core--jet, blazar-like interpretation viable. The spectral indices calculated between $8$ and $22$~GHz show that the shapes of the radio spectra of the middle and southern features are very similar. The FWHM size of the assumed CSO core, A3, is larger than that of A0, the southernmost component. Additionally, the CSO description of J1606$+$3124 does not conform straightforwardly to the standard picture of expanding radio galaxies \cite{Longair_1979}. According to that simple picture, assuming intrinsic symmetry, the approaching, brighter lobe should be seen farther away from the AGN center, while in J1606$+$3124, the arm-length ratio of the brighter to the fainter regions is smaller than unity. However, arm-length asymmetries in compact sources can also be produced by environmental gradients, so this argument is suggestive rather than decisive. Thus, our findings are also compatible with a core--jet scenario, where the center of the AGN is located at the brightest feature, A0, and a one-sided northern jet showing a slight wiggle at a few mas from the core can be detected. 

The relatively low value of $\delta$ was obtained assuming the equipartition value for the intrinsic brightness temperature. However, Homan et al. \cite{Homan2021} derived a lower median intrinsic brightness temperature, ${\sim}4 \cdot 10^{10}$\,K, using a large sample of quasars, implying \mbox{$\delta \sim 2$} for J1606$+$3124. Moreover, the $22$~GHz observing frequency corresponds to $124$~GHz in the rest frame of J1606$+$3124, and Cheng et al. \cite{Cheng2020} showed that the brightness temperature decreases with increasing frequency above $7$~GHz until $240$~GHz (in the source rest frame). This decrease is due to changes in the synchrotron opacity in the core region from optically thick to optically thin regimes. Thus, the low Doppler factor obtained at the $22$~GHz observing frequency does not by itself exclude the possibility that J1606$+$3124 is a mildly-beamed core--jet source.

In that picture, the polarization would originate from an ordered magnetic-field component in the core region. Because the absolute EVPA is not calibrated for J1606$+$3124, only the relative EVPA structure can in principle be discussed. Moreover, given the low fractional polarization, the detailed EVPA pattern should be treated cautiously. In the displayed, uncalibrated EVPA frame, the components A0 and A1 fitted to the total-intensity visibility data roughly coincide with regions where the plotted EVPAs differ by ${\sim}40^\circ$ in J1606$+$3124. For comparison, the highest-redshift blazar with detected polarized radio emission, J0906$+$6930, showed a polarized component at a projected distance of ${\sim}5$\,pc from the core at the similarly high rest-frame frequency of ${\sim}97$~GHz \cite{an20b}. This polarized feature was interpreted as arising due to jet--ISM interaction.

Multi-wavelength data of J1606$+$3124 indicate the presence of obscuration. X-ray measurements show a lack of ultraviolet and soft X-rays compared with quasars of similar redshift, and an obscuring column density of ${\gtrsim}10^{24}\;\mathrm{cm}^{-2}$, consistent with a Compton-thick absorber \cite{Xray2}. According to the colour index given by the difference between the $4.6$-$\upmu$m and $12$-$\upmu$m measurements \cite{Cutri} of the {\it Wide-field Infrared Survey Explorer} ({\it WISE}) space telescope, J1606$+$3124 is an obscured AGN \cite{Hickox}. The exact nature of J1606$+$3124 could be revealed if high-resolution optical data were available to pinpoint the location of the central engine. The source is not detected in the available data release of the \textit{Gaia} optical astrometry space mission, most likely because of its optical faintness ($R$ magnitude $22.56$~\cite{R_flux}), so no independent optical position of the nucleus with accuracy comparable to that of VLBI can presently be used for this purpose.

We note that the spectroscopic redshift of J1606$+$3124 is indicated to be uncertain by~\cite{R_flux}. The 8th data release of the Dark Energy Survey Instrument (DESI DR8 \cite{desi}) gives a photometric redshift estimate well below the spectroscopic value, $z_\mathrm{phot}=0.9\pm0.1$. By adopting this redshift, the brightness temperature of component A0 of J1606$+$3124 turns to be $2.5\cdot 10^{10}$\,K, lower than the equipartition limit, thus not requiring Doppler boosting of the radio emission. Additionally, the expansion speed derived from the changing separation of components N and S by \cite{an2022} using this lower redshift value would result in an apparent transverse speed of $(0.64 \pm 0.1)\,c$, and consequently a hot spot apparent advance speed of ${\sim}0.3\,c$. The projected linear extent of the radio-emitting structure would be ${\sim}71$\,pc if it were located at this lower redshift. All these properties are in accordance with the CSO classification of J1606$+$3124. Within this lower-redshift framework, the polarization detection at $22$\,GHz corresponds to a rest-frame frequency of ${\sim}42$\,GHz. Polarization was detected in the hot spot regions of a handful of CSOs in VLBI observations at $8$\,GHz and $15$\,GHz \cite{Gugliucci2007, Tremblay2016}. These showed that the closer, brighter side of the CSOs is polarized, possibly due to a lower level of Faraday depolarization. Thus, our detection of polarization can be interpreted in the CSO picture. However, we are not aware of the detection of a polarized signal in CSO hot spots at rest-frame frequencies as high as those in the case of J1606$+$3124.

\subsection{J1510+5702}

Our observation of J1510$+$5702 is in agreement with the blazar classification of this source, as we infer a high Doppler factor, $\delta \sim 21$, for the brightest core feature under the equipartition assumption. This object is one of the highest-redshift $\gamma$-ray emitters known. One $\gamma$-ray flare of J1510$+$5702 was detected on 4 February 2022 with a $0.1\text{--}300$-GeV flux ${\sim}25$ times larger than the value given in the fourth \textit{Fermi} Large Area Telescope catalog \cite{gokus2022}. 
 During the investigation of this flare, VLBI observations of J1510$+$5702 were performed at multiple frequencies ($15$, $22$, $43$, and $86$~GHz) in three epochs with an array consisting of the Very Long Baseline Array (VLBA), the Effelsberg radio telescope, and the Green Bank Telescope \cite{Benke}. The $22$-GHz radio structure of the source in our observation is similar to that found by \cite{Benke}, except that they detected the core and two jet components in the northern region in 2022 and 2023. One month after the $\gamma$-ray brightening, the jet component closest to the core had the highest flux density, which then continuously faded at $22$~GHz. Our observation took place $3.6$\,yr before the $\gamma$-ray flare (corresponding to ${\sim}0.7$\,yr in the source rest frame). At that time, the core component was the brightest feature. The elliptical shape of the core component, B0, and its PA towards the jet extension in our observation indicate that in 2018, the core and the closest jet feature seen by \cite{Benke} could not yet be resolved in our observation. This is further supported by the change in the core separation of this component derived by \cite{Benke}, which indicated a component ejection between 2016 and 2019. The core separation of B1 in our observation can be reconciled with the proper motion derived by \cite{Benke} for their C2 component. 

Polarization of J1510$+$5702 was detected with VLBI earlier, at $5$ and $8.4$~GHz \cite{OSullivan}. At $5$~GHz, polarization from both the core and the lobe feature could be detected, whereas at $8.4$~GHz, only the core was significantly polarized. At $22$~GHz, we similarly detected polarization from the core region only. The polarized flux density at $22$~GHz is $p_{22}\sim 3.4 \mathrm{\,mJy}$, comparable to the $8.4$-GHz value reported by \cite{OSullivan}, $p_{8.4}=(4.0\pm1.2)$\,mJy. 

The $22$~GHz observing frequency corresponds to $118$~GHz in the rest frame of J1510$+$5702. At such high frequencies, the emission is expected to arise farther upstream in the blazar jet than at cm wavelengths, where a more ordered magnetic field, less Faraday rotation, weaker depolarization, and consequently higher polarization degrees may occur (\cite{Agudo2014} and references therein). This was observed by \cite{Agudo2014} in simultaneous polarimetric observations of more than $200$ radio-loud AGNs at $86$ and $229$~GHz. They found median polarization degrees for flat-spectrum radio quasars of $m_{86}=3.2$\% and $m_{229}=7.7$\% at $86$ and $229$~GHz, respectively. We obtained a value similar to the former, ${\sim}3.5$\,\%, for the rest-frame $118$~GHz polarization degree for J1510$+$5702. Additionally, this value exceeds the polarization degree obtained by \cite{OSullivan} at $8$~GHz observing ($43$~GHz rest-frame) frequency, $m_8=(1.3\pm0.4)$\,\%. This is consistent with the finding of \cite{Agudo2014} that at mm wavelengths the degree of polarization increases with increasing frequencies, probably due to the lower Faraday depolarization effect, although the comparison here is based on non-simultaneous measurements. The diminished Faraday depolarization was also supported by the similar EVPAs they found at the two frequencies for their samples. 

The average EVPA in the core of J1510$+$5702 is $\chi_{22}\sim-73^\circ$ at $22$~GHz, while O'Sullivan et al. \cite{OSullivan} reported values of $\chi_{5}=-89^\circ \pm 3^\circ$ and $\chi_{8.4}=-93^\circ \pm 3^\circ$ at $5$ and $8.4$~GHz observing frequencies, respectively. The $22$-GHz EVPA is therefore broadly similar to the lower-frequency EVPAs within the limitations of the non-contemporaneous observations and the approximate absolute EVPA calibration. We do not derive an RM from these data. The comparison nevertheless does not contradict the expectation that Faraday rotation and depolarization become less important at higher emitted frequencies. This is similar to the findings of \cite{Agudo2014} for their samples. 

\section{Summary}

We conducted $22$-GHz EVN observations of the secure $z=4.31$ quasar J1510$+$5702 and the candidate high-redshift radio-loud AGN J1606$+$3124. If the published spectroscopic redshift value, $z=4.56$, of J1606$+$3124 is adopted, the $22$-GHz EVN observations probed rest-frame frequencies above $118$\,GHz. If J1606$+$3124 is instead located at $z_\mathrm{phot}\simeq0.9$, its emitted frequency is ${\sim}42$\,GHz. The radio brightness distributions of both sources are described by compact and extended Gaussian components on parsec scales.

{At $z=4.56$, the brightness temperature of component A0 of J1606$+$3124 lies modestly above the equipartition value and permits weak Doppler boosting; at $z_{\rm phot}=0.9$, it falls below equipartition. The $8.4\text{--}22$\,GHz comparison shows a robust spectral steepening toward the northern outer component, while the spectral indices of the southern and middle regions are indistinguishable. Low-level polarized signal was tentatively detected in component A0. None of these observables can uniquely distinguish between a mildly beamed core--jet geometry and a young compact symmetric source. A reliable redshift measurement and phase-referenced multi-epoch VLBI astrometry or an independent mas-scale optical/infrared localization of the central engine are therefore required to securely classify J1606$+$3124.}

Our VLBI results for J1510$+$5702 agree with its previous blazar classification and with later VLBI observations, showing a highly Doppler-boosted compact core and a southward-oriented jet. Its jet structure in 2018 is more compact than that observed more than $3$ years after our observation \cite{Benke}, consistent with the component motion reported by those authors. Its core is polarized with a higher degree of polarization than reported at lower frequencies~\cite{OSullivan}. The broadly similar EVPAs at observing frequencies from $5$ to $22$\,GHz are consistent with weaker Faraday effects at high rest-frame frequencies, although the present EVPA calibration does not allow a precise RM measurement.

Our study of {one $z>4$ quasar and one candidate high-redshift quasar} highlights how the rest-frame mm-wavelength total and polarized jet emission can be investigated at mas-scale resolution by employing relatively straightforward cm-wavelength VLBI experiments in the observer's frame. 

\vspace{6 pt} 

\authorcontributions{Conceptualization, T.A., S.F., J.Y., I.A., and Y.Z.; methodology, B.H. and K.\'E.G.; validation, K.\'E.G.; formal analysis, B.H. and K.\'E.G.; writing---original draft preparation, B.H. and K.\'E.G.; writing---review and editing, all authors; visualization, B.H. and K.\'E.G.; supervision, K.\'E.G. All authors have read and agreed to the published version of the manuscript.}

\funding{T.A. and Y.Z. are supported by the National SKA Program of China, grant number 2022SKA0120102. 
 This research was funded by the Hungarian National Research, Development and Innovation Office (NKFIH), under excellence grant TKP2021-NKTA-64. This project has received funding from the HUN-REN Hungarian Research Network.}

\dataavailability{\textls[-25]{The calibrated VLBI data are available from the corresponding author upon reasonable request. The raw VLBI data are publicly available from the EVN Data Archive at 
\url{https://doi.org/10.48717/1q5t-7508} and \url{https://doi.org/10.48717/2gyv-wb70}, accessed on 20 June 2026.}}

\acknowledgments{The European VLBI Network is a joint facility of independent European, African, Asian, and North American radio astronomy institutes. The scientific results presented in this publication are derived from data obtained under the following EVN project code: EZ028.}

\conflictsofinterest{The authors declare no conflicts of interest.}


\abbreviations{Abbreviations}{
The following abbreviations are used in this manuscript:\\

\noindent 
\begin{tabular}{@{}ll}
AGNs & active galactic nuclei \\
AIPS & Astronomical Image Processing System \\
CSO & compact symmetric object \\
EVN & European VLBI Network \\
EVPA & electric vector position angle \\
FSRQ & flat-spectrum radio quasar \\ 
FWHM & full width at half maximum \\
GPS & gigahertz-peaked spectrum \\
IF & intermediate frequency \\
ISM & interstellar medium \\
mas & milliarcsecond \\
MOJAVE & Monitoring of Jets in Active Galactic Nuclei with VLBA Experiments \\
NRAO & National Radio Astronomy Observatory \\
RM & rotation measure \\
rms & root mean square \\
SMBH & supermassive black hole \\
VLBA & very long baseline array \\
VLBI & very long baseline interferometry\\
WISE & Wide-field Infrared Survey Explorer\\
\end{tabular}
}




\begin{adjustwidth}{-\extralength}{0cm}

\reftitle{References}

\PublishersNote{}
\end{adjustwidth}
\end{document}